\documentclass[twocolumn]{aastex62}
\usepackage{xfrac}
\usepackage{appendix}

\newcommand{\gaia}{\textit{Gaia\/}}
\newcommand{\spitzer}{\textit{Spitzer\/}}
\newcommand{\wise}{\textit{WISE\/}}
\newcommand{\angstrom}{\textup{\AA}}
\newcommand{\RT}{radiative transfer}

\received{\today}
\revised{date}
\accepted{date}

\submitjournal{ApJ}

\shorttitle{Gaia 18dvy: a new FUor in the Cygnus OB3 association}
\shortauthors{Szegedi-Elek et al.}

\begin{document}

\title{Gaia 18dvy: a new FUor in the Cygnus OB3 association}


\author[0000-0001-7807-6644]{E. Szegedi-Elek}
\affil{Konkoly Observatory, Research Centre for Astronomy and Earth Sciences, H-1121 Budapest, Konkoly Thege \'ut 15--17, Hungary}

\author[0000-0001-6015-646X]{P. \'Abrah\'am}
\affil{Konkoly Observatory, Research Centre for Astronomy and Earth Sciences, H-1121 Budapest, Konkoly Thege \'ut 15--17, Hungary}
\affiliation{ELTE E\"otv\"os Lor\'and University, Institute of Physics, P\'azm\'any P\'eter s\'et\'any 1/A, 1117 Budapest, Hungary}

\author[0000-0002-9658-6151]{{\L}. Wyrzykowski}
\affil{Warsaw University Astronomical Observatory, Al. Ujazdowskie 4, 00-478 Warszawa, Poland}

\author[0000-0002-7538-5166]{M. Kun}
\affil{Konkoly Observatory, Research Centre for Astronomy and Earth Sciences, H-1121 Budapest, Konkoly Thege \'ut 15--17, Hungary}

\author[0000-0001-7157-6275]{\'A. K\'osp\'al}
\affil{Konkoly Observatory, Research Centre for Astronomy and Earth Sciences, H-1121 Budapest, Konkoly Thege \'ut 15--17, Hungary}
\affil{Max Planck Institute for Astronomy, K\"onigstuhl 17, D-69117 Heidelberg, Germany}
\affiliation{ELTE E\"otv\"os Lor\'and University, Institute of Physics, P\'azm\'any P\'eter s\'et\'any 1/A, 1117 Budapest, Hungary}

\author[0000-0003-2835-1729]{L. Chen}
\affil{Konkoly Observatory, Research Centre for Astronomy and Earth Sciences, H-1121 Budapest, Konkoly Thege \'ut 15--17, Hungary}

\author[0000-0002-1326-1686]{G. Marton}
\affil{Konkoly Observatory, Research Centre for Astronomy and Earth Sciences, H-1121 Budapest, Konkoly Thege \'ut 15--17, Hungary}
\affiliation{ELTE E\"otv\"os Lor\'and University, Institute of Physics, P\'azm\'any P\'eter s\'et\'any 1/A, 1117 Budapest, Hungary}

\author{A. Mo\'or}
\affil{Konkoly Observatory, Research Centre for Astronomy and Earth Sciences, H-1121 Budapest, Konkoly Thege \'ut 15--17, Hungary}
\affiliation{ELTE E\"otv\"os Lor\'and University, Institute of Physics, P\'azm\'any P\'eter s\'et\'any 1/A, 1117 Budapest, Hungary}

\author[0000-0002-8722-6875]{Cs. Kiss}
\affil{Konkoly Observatory, Research Centre for Astronomy and Earth Sciences, H-1121 Budapest, Konkoly Thege \'ut 15--17, Hungary}
\affiliation{ELTE E\"otv\"os Lor\'and University, Institute of Physics, P\'azm\'any P\'eter s\'et\'any 1/A, 1117 Budapest, Hungary}

\author[0000-0001-5449-2467]{A. P\'al}
\affil{Konkoly Observatory, Research Centre for Astronomy and Earth Sciences, H-1121 Budapest, Konkoly Thege \'ut 15--17, Hungary}
\affil{Department of Astronomy, Lor\'and E\"otv\"os University, P\'azm\'any P. s\'et\'any. 1/A, H-1117 Budapest, Hungary}
  \affiliation{ELTE E\"otv\"os Lor\'and University, Institute of Physics, P\'azm\'any P\'eter s\'et\'any 1/A, 1117 Budapest, Hungary}

\author[0000-0002-2046-4131]{L. Szabados}
\affil{Konkoly Observatory, Research Centre for Astronomy and Earth Sciences, H-1121 Budapest, Konkoly Thege \'ut 15--17, Hungary}

\author[0000-0003-4989-575X]{J. Varga}
\affil{Leiden Observatory, Leiden University, PO Box 9513, NL2300, RA Leiden, The Netherlands}
\affil{Konkoly Observatory, Research Centre for Astronomy and Earth Sciences, H-1121 Budapest, Konkoly Thege \'ut 15--17, Hungary}

\author[0000-0002-2655-5069]{E. Varga-Vereb\'elyi}
\affil{Konkoly Observatory, Research Centre for Astronomy and Earth Sciences, H-1121 Budapest, Konkoly Thege \'ut 15--17, Hungary}

\author{C. Andreas}
\affil{Astrophysikalisches Institut und Universitäts-Sternwarte, FSU Jena,
Schillergäßchen 2-3, D-07745 Jena, Germany}

\author[0000-0002-6578-5078]{E. Bachelet}
\affil{Las Cumbres Observatory, 6740 Cortona Drive, Suite 102,93117 Goleta, CA, USA}

\author{R. Bischoff}
\affil{Astrophysikalisches Institut und Universitäts-Sternwarte, FSU Jena,
Schillergäßchen 2-3, D-07745 Jena, Germany}

\author[0000-0002-8585-4544]{A. B\'odi}
\affil{Konkoly Observatory, Research Centre for Astronomy and Earth Sciences, H-1121 Budapest, Konkoly Thege \'ut 15--17, Hungary}
\affil{MTA CSFK Lend\"ulet Near-Field Cosmology Research Group, Konkoly Thege Mikl\'os \'ut 15-17, H-1121 Budapest, Hungary}

\author[0000-0001-6180-3438]{E. Breedt}
\affil{Institute of Astronomy, University of Cambridge, Madingley Road, Cambridge, CB3 0HA, UK}

\author[0000-0003-0126-3999]{U. Burgaz}
\affil{Department of Astronomy, Kyoto University, Kitashirakawa-Oiwakecho, Sakyo-ku, Kyoto 606-8502, Japan}
\affil{Department of Astronomy and Space Sciences, Ege University, 35100, Izmir, Turkey}

\author[0000-0002-2853-0834]{T. Butterley}
\affil{Centre for Advanced Instrumentation, Durham University, UK}

\author{J. M. Carrasco}
\affil{Dept. Física Qu\'antica i Astrofísica, Institut de Ci\'encies del Cosmos (ICCUB), Universitat de Barcelona (IEEC-UB), Mart\'i Franqu\'es 1, E08028 Barcelona, Spain}

\author{V. \v{C}epas}
\affil{Institute of Theoretical Physics and Astronomy, Vilnius University, Saul\.etekio av. 3, 10257 Vilnius, Lithuania}

\author[0000-0002-6710-6868]{G. Damljanovic}
\affil{Astronomical Observatory, Volgina 7, 11060 Belgrade, Serbia}

\author{I. Gezer}
\affil{Warsaw University Astronomical Observatory, Al. Ujazdowskie 4, 00-478 Warszawa, Poland}

\author{V. Godunova}
\affil{ICAMER Observatory, NAS of Ukraine, 27 Acad. Zabolotnoho Str., 03143, Kyiv, Ukraine}

\author[0000-0002-1650-1518]{M. Gromadzki}
\affil{Warsaw University Astronomical Observatory, Al. Ujazdowskie 4, 00-478 Warszawa, Poland}

\author{A. Gurgul}
\affil{Warsaw University Astronomical Observatory, Al. Ujazdowskie 4, 00-478 Warszawa, Poland}

\author{L. Hardy}
\affil{Department of Physics and Astronomy, University of Sheffield, Sheffield, S3 7RH, United Kingdom}

\author{F. Hildebrandt}
\affil{Astrophysikalisches Institut und Universitäts-Sternwarte, FSU Jena,
Schillergäßchen 2-3, D-07745 Jena, Germany}

\author{S. Hoffmann}
\affil{Astrophysikalisches Institut und Universitäts-Sternwarte, FSU Jena,
Schillergäßchen 2-3, D-07745 Jena, Germany}

\author[0000-0003-0961-5231]{M. Hundertmark}
\affil{Zentrum f{\"u}r Astronomie der Universit{\"a}t Heidelberg, Astronomisches Rechen-Institut, M{\"o}nchhofstr. 12-14, 69120 Heidelberg, Germany}

\author{N. Ihanec}
\affil{Warsaw University Astronomical Observatory, Al. Ujazdowskie 4, 00-478 Warszawa, Poland}

\author{R. Janulis}
\affil{Institute of Theoretical Physics and Astronomy, Vilnius University, Saul\.etekio av. 3, 10257 Vilnius, Lithuania}

\author{Cs. Kalup}
\affil{Konkoly Observatory, Research Centre for Astronomy and Earth Sciences, H-1121 Budapest, Konkoly Thege \'ut 15--17, Hungary}

\author{Z. Kaczmarek}
\affil{Warsaw University Astronomical Observatory, Al. Ujazdowskie 4, 00-478 Warszawa, Poland}

\author{R. K\"onyves-T\'oth}
\affil{Konkoly Observatory, Research Centre for Astronomy and Earth Sciences, H-1121 Budapest, Konkoly Thege \'ut 15--17, Hungary}

\author[0000-0002-8813-4884]{M. Krezinger}
\affil{Konkoly Observatory, Research Centre for Astronomy and Earth Sciences, H-1121 Budapest, Konkoly Thege \'ut 15--17, Hungary}

\author[0000-0002-2729-5369]{K. Kruszy{\'n}ska}
\affil{Warsaw University Astronomical Observatory, Al. Ujazdowskie 4, 00-478 Warszawa, Poland}

\author[0000-0001-7221-855X]{S. Littlefair}
\affil{Department of Physics and Astronomy, University of Sheffield, Sheffield, S3 7RH, United Kingdom}

\author{M. Maskoli{\=u}nas}
\affil{Institute of Theoretical Physics and Astronomy, Vilnius University, Saul\.etekio av. 3, 10257 Vilnius, Lithuania}

\author{L. M\'esz\'aros}
\affil{Konkoly Observatory, Research Centre for Astronomy and Earth Sciences, H-1121 Budapest, Konkoly Thege \'ut 15--17, Hungary}

\author[0000-0001-8916-8050]{P. Miko{\l}ajczyk}
\affil{Astronomical Institute, University of Wroc{\l}aw, ul. Kopernika 11, 51-622 Wroc{\l}aw, Poland}

\author{M. Mugrauer}
\affil{Astrophysikalisches Institut und Universitäts-Sternwarte, FSU Jena,
Schillergäßchen 2-3, D-07745 Jena, Germany}

\author{H. Netzel}
\affil{Nicolaus Copernicus Centre of Polish Academy of Sciences, ul. Bartycka 18, 00-716 Warszawa, Poland}

\author{A. Ordasi}
\affil{Konkoly Observatory, Research Centre for Astronomy and Earth Sciences, H-1121 Budapest, Konkoly Thege \'ut 15--17, Hungary}

\author[0000-0002-3326-2918]{E. Pak{\v s}tien{\.e}}
\affil{Institute of Theoretical Physics and Astronomy, Vilnius University, Saul\.etekio av. 3, 10257 Vilnius, Lithuania}

\author[0000-0002-9326-9329]{K. A. Rybicki}
\affil{Warsaw University Astronomical Observatory, Al. Ujazdowskie 4, 00-478 Warszawa, Poland}

\author[0000-0003-0926-3950]{K. S\'arneczky}
\affil{Konkoly Observatory, Research Centre for Astronomy and Earth Sciences, H-1121 Budapest, Konkoly Thege \'ut 15--17, Hungary}

\author{B. Seli}
\affil{Konkoly Observatory, Research Centre for Astronomy and Earth Sciences, H-1121 Budapest, Konkoly Thege \'ut 15--17, Hungary}

\author{A. Simon}
\affil{Taras Shevchenko National University of Kyiv,
Glushkova ave., 4, 03127, Kyiv, Ukraine}

\author{K. \v{S}i\v{s}kauskait\.e}
\affil{Institute of Theoretical Physics and Astronomy, Vilnius University, Saul\.etekio av. 3, 10257 Vilnius, Lithuania}

\author[0000-0001-7806-2883]{\'A. S\'odor}
\affil{Konkoly Observatory, Research Centre for Astronomy and Earth Sciences, H-1121 Budapest, Konkoly Thege \'ut 15--17, Hungary}

\author[0000-0001-5991-6863]{K. V. Sokolovsky}
\affil{Department of Physics and Astronomy, Michigan State University, 567 Wilson Rd, East Lansing, MI 48824, USA}
\affil{Sternberg Astronomical Institute, Moscow State University, Universitetskii~pr.~13, 119992~Moscow, Russia}
\affil{Astro Space Center of Lebedev Physical Institute, Profsoyuznaya~St.~84/32, 117997~Moscow, Russia}

\author{W. Stenglein}
\affil{Astrophysikalisches Institut und Universitäts-Sternwarte, FSU Jena,
Schillergäßchen 2-3, D-07745 Jena, Germany}
\author{R. Street}
\affil{Las Cumbres Observatory, 6740 Cortona Drive, Suite 102,93117 Goleta, CA, USA} 

\author[0000-0002-1698-605X]{R. Szak\'ats}
\affil{Konkoly Observatory, Research Centre for Astronomy and Earth Sciences, H-1121 Budapest, Konkoly Thege \'ut 15--17, Hungary}

\author[0000-0002-3697-2616]{L. Tomasella}
\affil{INAF Osservatorio Astronomico di Padova, Vicolo dell'Osservatorio 5, 35122 Padova, Italy}

\author{Y. Tsapras}
\affil{Zentrum f{\"u}r Astronomie der Universit{\"a}t Heidelberg, Astronomisches Rechen-Institut, M{\"o}nchhofstr. 12-14, 69120 Heidelberg, Germany}

\author[0000-0002-6471-8607]{K. Vida}
\affil{Konkoly Observatory, Research Centre for Astronomy and Earth Sciences, H-1121 Budapest, Konkoly Thege \'ut 15--17, Hungary}
\affiliation{ELTE E\"otv\"os Lor\'and University, Institute of Physics, P\'azm\'any P\'eter s\'et\'any 1/A, 1117 Budapest, Hungary}

\author{J. Zdanavi\v{c}ius}
\affil{Institute of Theoretical Physics and Astronomy, Vilnius University, Saul\.etekio av. 3, 10257 Vilnius, Lithuania}

\author{M. Zieli{\'n}ski}
\affil{Warsaw University Astronomical Observatory, Al. Ujazdowskie 4, 00-478 Warszawa, Poland}

\author{P. Zieli{\'n}ski}
\affil{Warsaw University Astronomical Observatory, Al. Ujazdowskie 4, 00-478 Warszawa, Poland}

\author{O. Zi{\'o}{\l}kowska}
\affil{Warsaw University Astronomical Observatory, Al. Ujazdowskie 4, 00-478 Warszawa, Poland}

\begin{abstract}
  We present optical-infrared photometric and spectroscopic observations of Gaia\,18dvy, located in the Cygnus~OB3 association at a distance of 1.88\,kpc. The object was noted by the Gaia alerts system when its lightcurve exhibited a $\gtrsim$4~mag rise in 2018-2019. The brightening was also observable at mid-infared wavelengths. The infrared colors of Gaia\,18dvy became bluer as the outburst progressed. Its optical and near-infrared spectroscopic characteristics in the outburst phase are consistent with those of bona fide FU Orionis-type young eruptive stars. The progenitor of the outburst is probably a low-mass K-type star with an optical extinction of $\sim$3\,mag. A radiative transfer modeling of the circumstellar structure, based on the quiescent spectral energy distribution, indicates a disk with a mass of $4{\times}10^{-3}\,M_{\odot}$. Our simple accretion disk modeling implies that the accretion rate had been exponentially increasing for more than 3 years until mid-2019, when it reached a peak value of $6.9 \times 10^{-6}\,M_{\odot}$yr$^{-1}$. In many respects, Gaia\,18dvy is similar to the FU~Ori-type object HBC~722.
\end{abstract}




\section{Introduction}
\label{sec:intro}

FU~Orionis-type young eruptive stars (FUors) form a small but important subclass of Sun-like pre-main-sequence stars. They exhibit a brightening of up to 5\,mag during several months or years, followed by a fading phase of several decades or a century \citep{herbig77,hk96,audard2014}. Their outbursts are powered by enhanced accretion from the circumstellar disk onto the star. FUors are often surrounded by thick envelopes, drive jets and outflows, and exhibit a characteristic absorption spectrum \citep{connelley2018}. 

If all Sun-like young stars undergo eruptive phases, then a sizeable part of their final stellar mass may build up during repeated outbursts \citep[e.g.][]{vorobyov2006}, and characterizing the FUor phenomenon would be fundamental to understand the formation of low-mass stars. The physical origin of the enhanced accretion is still debated: thermal instability, combination of gravitational and magnetorotational instabilities, disk fragmentation and environmental triggers are invoked \citep[for a review see][]{audard2014}. To decide between these scenarios, a larger sample of FUors needs to be analysed, however, their known population is still very small: \citet{audard2014} listed only 26 FUors and FUor-like objects. Therefore any new discovery may provide important insights into the physics of episodic accretion.

The {\gaia} Photometric Science Alerts System \citep{wyrzykowski2012, hodgkin2013} contributes to the field of star and planet formation by discovering and publishing otherwise unnoticed brightenings and fadings of young stellar objects. Up to now, two alerts were proven to be  young eruptive stars: Gaia\,17bpi \citep{hillenbrand2018}, and Gaia\,19ajj \citep{hillenbrand2019}.

In this paper we present a detailed analysis of Gaia\,18dvy\footnote{ http://gsaweb.ast.cam.ac.uk/alerts/alert/Gaia18dvy/} (RA$_{\rm J2000}$ = 20$^{\rm h}$05$^{\rm m}$06$.\!\!^{\rm s}$02, Dec$_{\rm J2000}$ = +36$^{\circ}$29$'$13$\farcs$5, ID: {\it Gaia}  DR2 2059895933266183936), a {\gaia} alert source whose ${\gtrsim}4$ mag brightness increase was published on 2018 December 19. The timescale and amplitude of the brightening suggested a FUor outburst. We carried out optical photometric monitoring of the source, and obtained  optical and infrared spectra. Here we combine these with archival optical, near- and mid-infrared data, and apply simple models to understand the nature of the object and the brightening process.


\section{Observations and data reduction}
\label{sec:obs}


\subsection{Photometry}

\begin{figure*}
   \includegraphics[width=\textwidth]{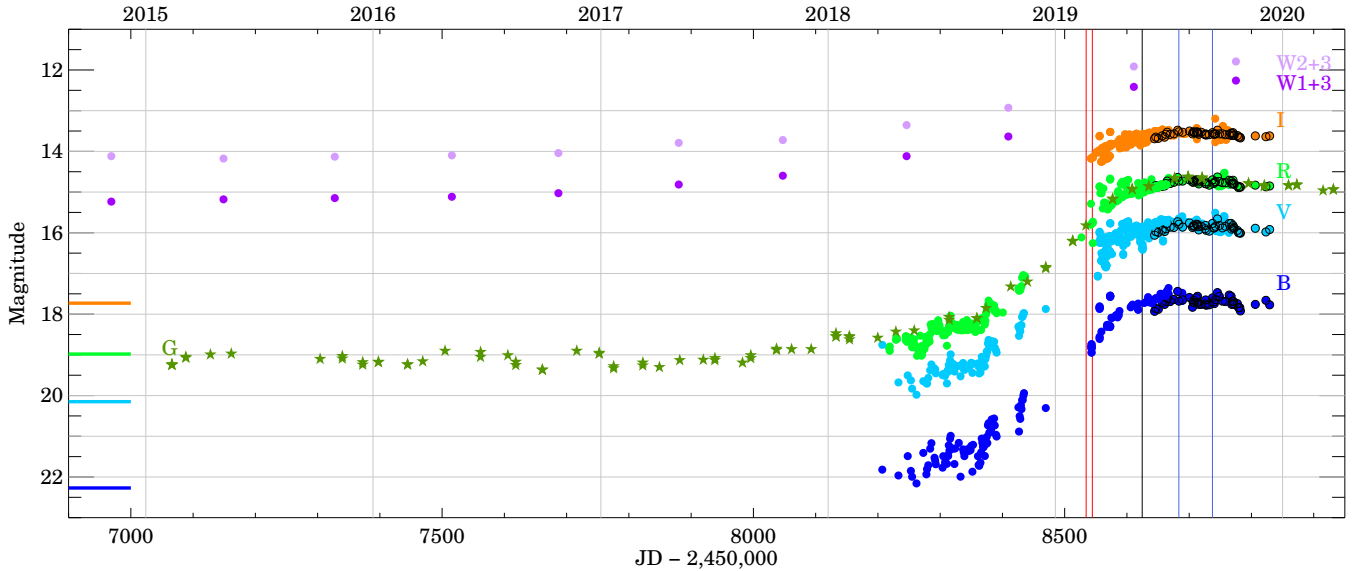}\\
  \caption{Optical and infrared light curves of Gaia\,18dvy. Green asterisks show {\it Gaia} data, purple dots show {\wise} data, filled dots indicate ZTF (converted to the Johnson-Cousins system) and OPTICON data, while our photometry from the Konkoly Observatory is highlighted by black circles. Average Pan-STARRS magnitudes, converted to the Johnson-Cousins system, are indicated by the horizontal lines at the left side of the figure. Red vertical lines mark when we took optical spectra of Gaia\,18dvy, while the black vertical line indicates the epoch of our NIR spectrum. The two blue vertical lines display the time period when the TESS satellite observed Gaia\,18dvy. Follow-up photometric data are available in Table \ref{follow}.
  }\label{fig:light}
\end{figure*}

We downloaded multi-epoch {\gaia} $G$-band photometry for Gaia\,18dvy from the alerts service webpage and plotted the light curve in Fig.~\ref{fig:light}. We supplemented these with data available in public databases and with our own new observations.

\begin{figure}
\includegraphics[width=\columnwidth]{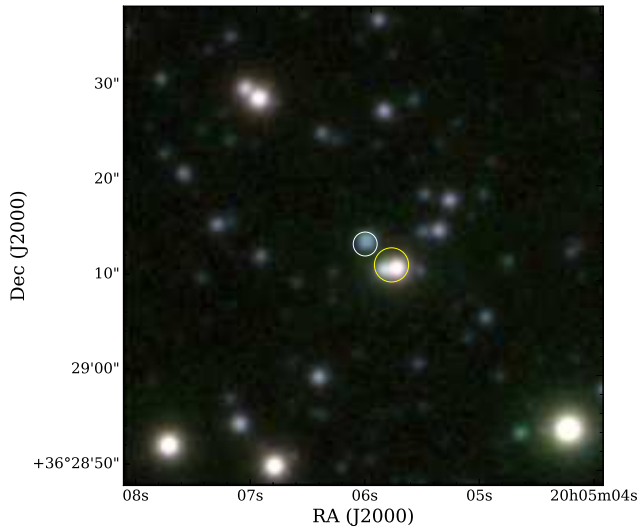}
    \caption{False color composite image centered on Gaia\,18dvy (white circle) using Pan-STARRS $i,z,y$ images. The nearby sources whose contribution was subtracted from the WISE photometry are marked by the yellow circle.}\label{fig:env}
\end{figure}

The Pan-STARRS \citep{panstarrs} survey provided light curves for Gaia\,18dvy in $grizy$ filters between 2009 July and 2014 June. According to the epoch photometry, the source was constant during this period to within 0.1-0.3\,mag, therefore we only plot the mean magnitudes in Fig.~\ref{fig:light} to indicate the quiescent brightness levels, after we converted the Sloan magnitudes to Johnson--Cousins magnitudes using equations from \citet{pan}. In Fig.~\ref{fig:env} we show the environment of Gaia\,18dvy using Pan-STARRS images.

Gaia\,18dvy was covered by the Zwicky Transient Facility (ZTF, \citealt{ztf}), a new time-domain survey at Palomar Observatory in operation since 2018 February. We downloaded $g$ and $r$ band photometry from the second data release from the NASA/IPAC Infrared Science Archive (IRSA), which contains data until June 2019. There are no specific conversion formulae for the ZTF filters, therefore we converted the ZTF magnitudes to the Johnson--Cousins system using the equations of  \citet{pan},
considering that the ZTF filter profiles are not very different from the Sloan filters. We plotted the resulting $BVR\mathrm{\sb{C}}$ light curves in Fig.~\ref{fig:light}.

We observed Gaia\,18dvy in the $BVR\mathrm{\sb{C}}I\mathrm{\sb{C}}$ bands between 2019 June and December using the 60/90/180\,cm Schmidt telescope at the Konkoly Observatory (Hungary). Because Gaia\,18dvy has two nearby stars within $\sim$4$''$ (marked in Fig.~\ref{fig:env}), we performed aperture photometry with a small aperture radius of 2$''$ to minimize contamination. We transformed the instrumental magnitudes to the standard system using comparison stars from the Pan-STARRS catalog \citep{panstarrs}, after transforming the Pan-STARRS magnitudes to the Johnson--Cousins system as before. These results, highlighted with circles, are also plotted in Fig.~\ref{fig:light}.


We monitored Gaia\,18dvy at optical wavelengths using the OPTICON Time-Domain Follow-up Network\footnote{The OPTICON Time-Domain   Follow-up Network includes the following telescopes: pt5m telescope at the Roque de los Muchachos Observatory on La Palma \citep{hardy2015}; 0.8\,m Telescopi Joan Oro (TJO) at l'Observatori Astronomic del Montsec in Spain; 1.4\,m telescope at the Astronomical Station Vidojevica, near Prokuplje, Serbia; 0.6\,m Bia\l{}k\'ow Observatory, operated by the Astronomical Institute of the University of Wroc\l{}aw, Poland; 0.35\,m Cassegrain and 1.65\,m Ritchey--Chretien telescopes of Mol\.etai Astronomical Observatory in Mol\.etai, Kulionys, Lithuania; 2.3\,m Aristarchos Telescope at Helmos Observatory, Peloponnese, Greece; 2\,m Ritchey-Chretien and 0.6\,m Cassegrain telescopes at the Terskol Observatory (the North Caucasus, Russia) operated by ICAMER of NAS of Ukraine; 0.6\,m Ritchey-Chretien telescope of the Michigan State University Observatory (MPC code 766), USA.} since 2019 February. All follow-up images were standardized in an automated fashion by the Cambridge Photometric Calibration Server \citep[CPCS,][]{zielinski2019}. To account for differences in filters, comparison stars, and aperture size, we shifted the photometry obtained by the OPTICON network telescopes to match with our Konkoly Schmidt data.

Gaia\,18dvy was also monitored with the Las Cumbres Observatory network of robotic telescopes \citep{brown2013}. About 200 images have been obtained in $V$ and $I\mathrm{\sb{C}}$ and automatically reduced using the BANZAI pipeline \citep{mccully2019}. Similary to the OPTICON data, photometry and calibration has been obtained using the CPCS pipeline.

Gaia\,18dvy was observed with the Schmidt-Teleskop-Kamera \citep[STK,][]{jena} of University Observatory Jena in the  Bessell V,R,I-bands. Each night two frames (60 sec) were taken in each filter.  Standard data reduction was performed with dark frames and sky- or domeflats taken in each night before or after the observations in twilight.

\begin{figure}
   \includegraphics[width=\columnwidth]{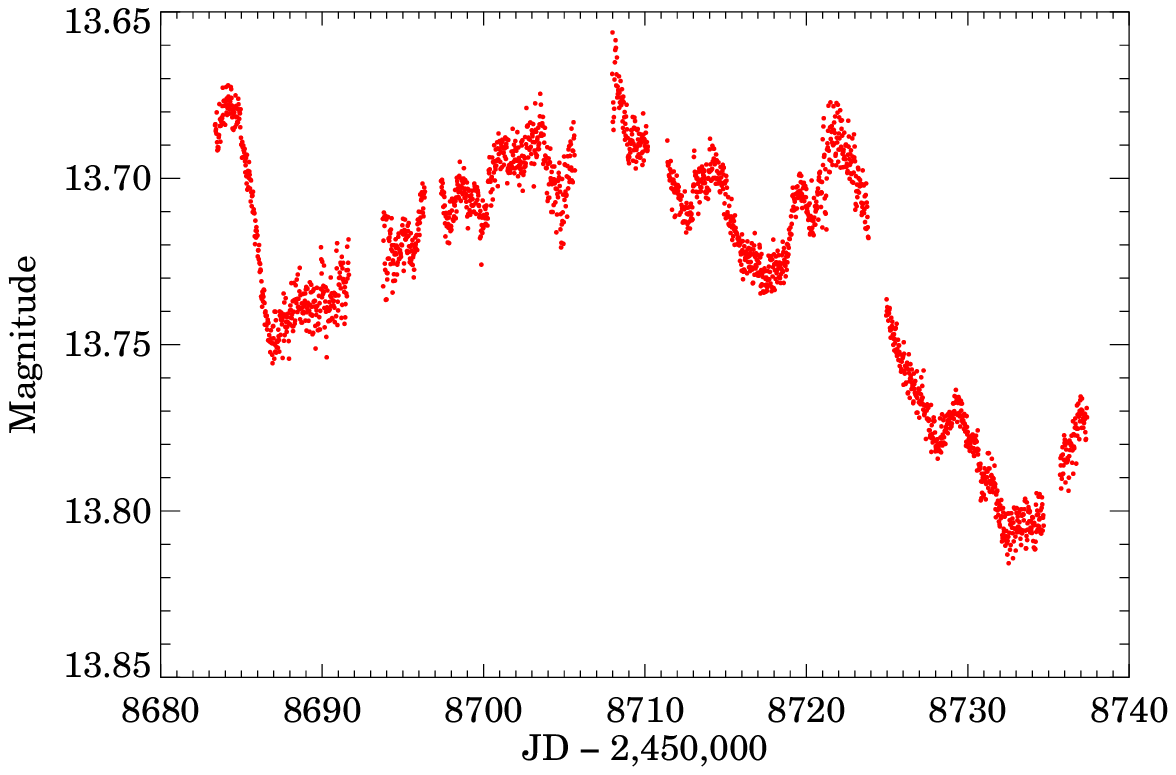}
   \includegraphics[width=\columnwidth]{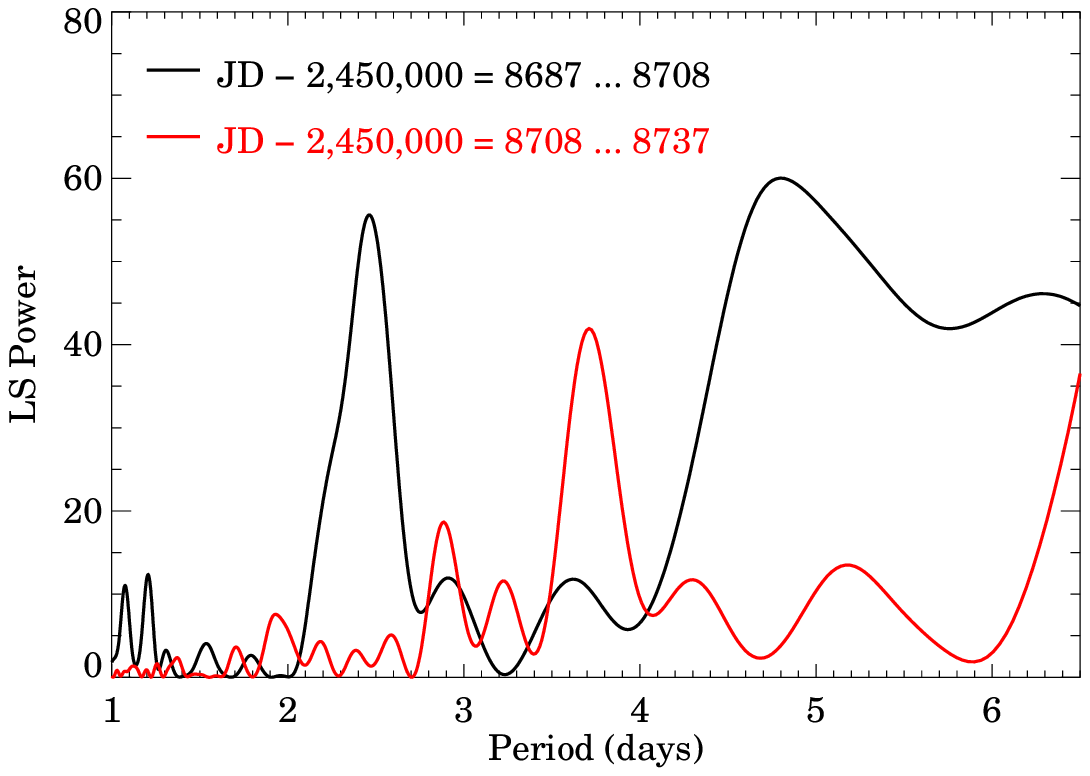}
   \caption{Top: {\it TESS} light curve of Gaia\,18dvy. Bottom: Lomb--Scargle periodogram of different parts of the TESS light curve after the subraction of a linear trend.}\label{fig:tess}
\end{figure}

Gaia\,18dvy was observed by the Transiting Exoplanet Survey Satellite \citep[TESS,][]{ricker2015} during Sectors 14 and 15 (2019 July 18 to 2019 September 10). We retrieved the full-frame images from the MAST archive and analyzed using a FITSH-based pipeline \citep{pal2012} providing convolution-based differential imaging algorithms and subsequent photometry on the residual images. Because the spectral sensitivity of the TESS detectors are close to the $I\mathrm{\sb{C}}$-band filter, we used our contemporaneous Schmidt $I\mathrm{\sb{C}}$-band data for the absolute calibration of the TESS photometry. The resulting light curve is shown in Fig.~\ref{fig:tess}.

We obtained $JHK_{\rm S}$ images of Gaia\,18dvy on 2019 July 4 using the Wide Field Camera of the NOTCam instrument on the Nordic Optical Telescope (La Palma, Spain). The instrumental magnitudes, obtained by aperture photometry, were calibrated using 2MASS magnitudes of bright comparison stars in the field of view. In the $K_{\rm S}$ band the source was already in the nonlinear regime of the detector. To correct for this, we determined an empirical relation based on a set of stars comparable in brightness to Gaia\,18dvy, similarly to \citet{kospal2017}. The results are $J=11.25\pm0.02$\,mag, $H=10.36\pm0.03$\,mag, and $K_{\rm S}=9.7\pm0.1$\,mag, indicating significant brightening compared to photometry similarly obtained in UKIDSS \citep{lawrence2007} images from 2009 August ($J=15.73\pm0.06$\,mag, $H=14.68\pm0.07$\,mag, $K_{\rm S}=13.70\pm0.08$\,mag).


Gaia\,18dvy was monitored with a twice-yearly cadence by the {\it Wide-field Infrared Survey Explorer} ({\wise}, \citealt{wise_ref}) in the W1 (3.4 $\micron$) and W2 (4.6 $\micron$) bands between 2015 and 2019, as part of the {\it NEOWISE} Reactivation project. For each epoch, we downloaded time resolved observations from the {\it NEOWISE-R} Single Exposure Source Table and computed their seasonal averages after removing outlier points. Since the beam size of {\wise} is ${\sim}6''$ in these bands, contamination from the neighbouring sources (Fig.~\ref{fig:env}) had to be taken into account. We used {\spitzer} IRAC fluxes of these sources from the GLIMPSE360 catalog at IRSA \citep{whitney2011} and subtracted 1.65\,mJy at 3.6$\,\micron$ and 1.08\,mJy at 4.5$\,\micron$ from the WISE fluxes of Gaia\,18dvy, assuming that the measured fluxes would be very similar in the Spitzer and WISE systems, and that the neighboring sources were constant in time. 

\begin{deluxetable}{llll}
 \tablecaption{Follow-up photometry \label{follow}}
 
 \tablehead{
 \colhead{MJD } &\colhead{Filter}& \colhead{Magnitude} &\colhead{Instrument}\\ 
 }
 \startdata 
8756.384 &i &13.54  $\pm$ 0.08 &ptm5    \\
8757.376 &V &15.94  $\pm$  0.05 &ptm5    \\
8757.380 &r &14.79  $\pm$  0.08 &ptm5    \\
8757.384 &i &13.69  $\pm$  0.07 &ptm5    \\
8758.343 &B &17.69  $\pm$  0.03 &Konkoly Schmidt   \\
8758.343 &V &15.84  $\pm$  0.02 &Konkoly Schmidt   \\
8758.343 &R &14.77  $\pm$  0.01 &Konkoly Schmidt  \\
8758.343 &I& 13.59  $\pm$  0.01& Konkoly Schmidt\\  
\enddata
 \tablecomments{ This table is available in its entirety in a machine-readable form in the online journal. A portion is shown here for guidance regarding its form and content.}
\end{deluxetable}

\begin{figure*}
   \includegraphics[width=0.492\textwidth]{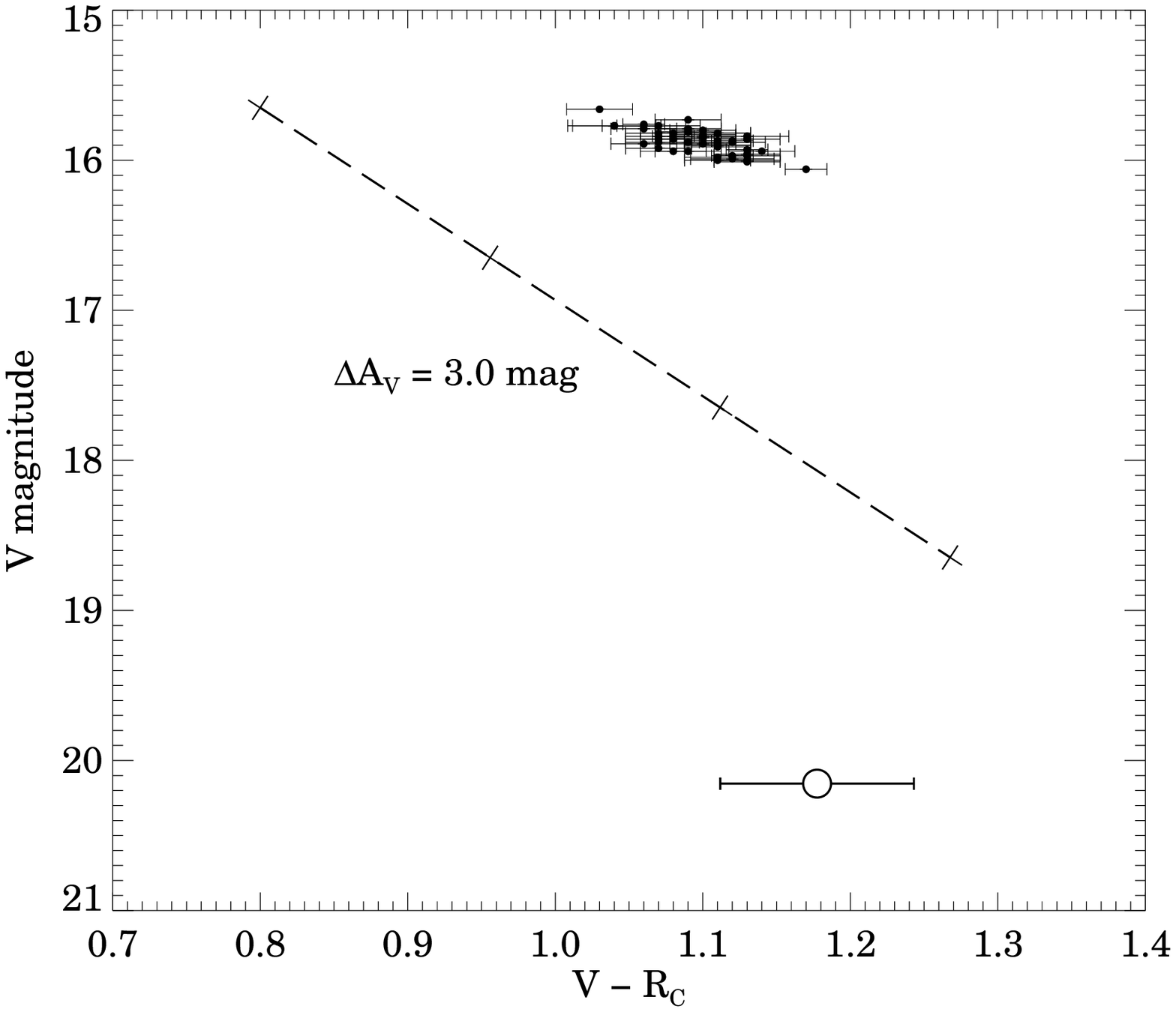}
   \includegraphics[width=0.47\textwidth]{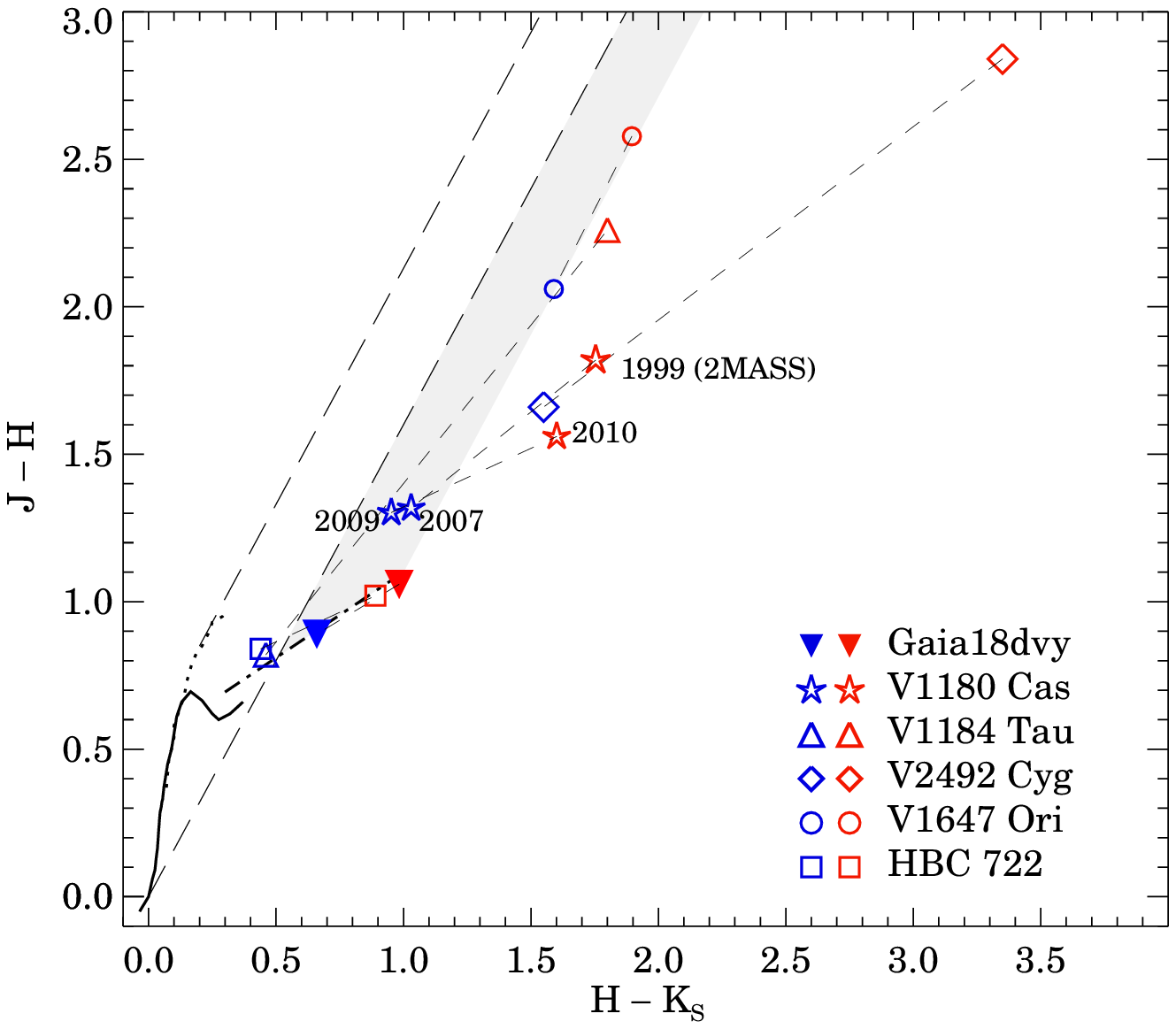}
   \caption{Left: Optical color-magnitude diagram. Filled symbols are observations obtained later than 2019 June with the Schmidt telescope at Konkoly Observatory, Hungary. Empty circle corresponds to the pre-outburst values based on Pan-STARRS. 
   The dashed line is the $R_V$=3.1 extinction path from $A_V$=15.5 to $A_V$=18.5 mag.
   Right:
   $J - H$ vs.~$H - K_{\rm S}$ color-color diagram. The solid curve indicates the zero-age main sequence, the long-dashed lines show the reddening path \citep{cardelli89}. The dash-dotted line is the locus of unreddened T~Tauri stars \citep{meyer97}, and the gray  band indicates the area occupied by reddened pre-main-sequence stars. For comparison, color variations of V1647~Ori \citep{acosta2007}, V1180~Cas \citep{kun2011}, V1184~Tau \citep{grinin2009}, HBC~722, and V2492~Cyg \citep{kospal2011} are indicated by blue (outburst) and red (quiescence) symbols.}\label{fig:tcd}
\end{figure*}

\subsection{Spectroscopy}

We obtained an optical spectrum of Gaia\,18dvy with the Isaac Newton Telescope  (La Palma, Spain) on 2019 February 20, using the Intermediate Dispersion Spectrograph fitted with the R300V grating, which covered the $345-800$\,nm range, and gave $R \sim 1000$ resolution with the 1$\arcsec$ slit. The exposure time was 600\,s. The spectrum was reduced and calibrated using the {\sc Starlink} suite of tools. The wavelength solution was derived from Copper-Neon and Copper-Argon arc lamp exposures.

We took an optical spectrum  on 2019 February 28 at the Copernico 1.82\,m telescope operated by INAF-Osservatorio Astronomico di Padova (Asiago, Italy), using the Asiago Faint Object Spectrograph AFOSC). We acquired spectroscopy with the VPH6 (450--1000\,nm, $R \sim 500$) and VPH7 (320--700\,nm, $R \sim 470$) grisms and the 1$\farcs69$ slit. The exposure time was 2$\times$1200\,s. The extracted spectra were wavelength-calibrated using comparison lamp spectra and flux-calibrated using spectrophotometric standard stars Feige~66 and BD+33~2642. Telluric absorption was corrected using the spectra of both telluric and spectrophotometric standards.

We obtained a near-infrared (NIR) spectrum of Gaia\,18dvy on 2019 May 21 with NOTCam using the 0$\farcs$6 slit, which provided a resolution of $R \sim 2500$. The total exposure time was 1280\,s. Spectra of Xenon and Argon lamps were observed for wavelength calibration, and a halogen lamp for flatfielding. The O9.5IV-type star HD~192001 was observed for telluric correction. 

\begin{figure}
\includegraphics[width=\columnwidth]{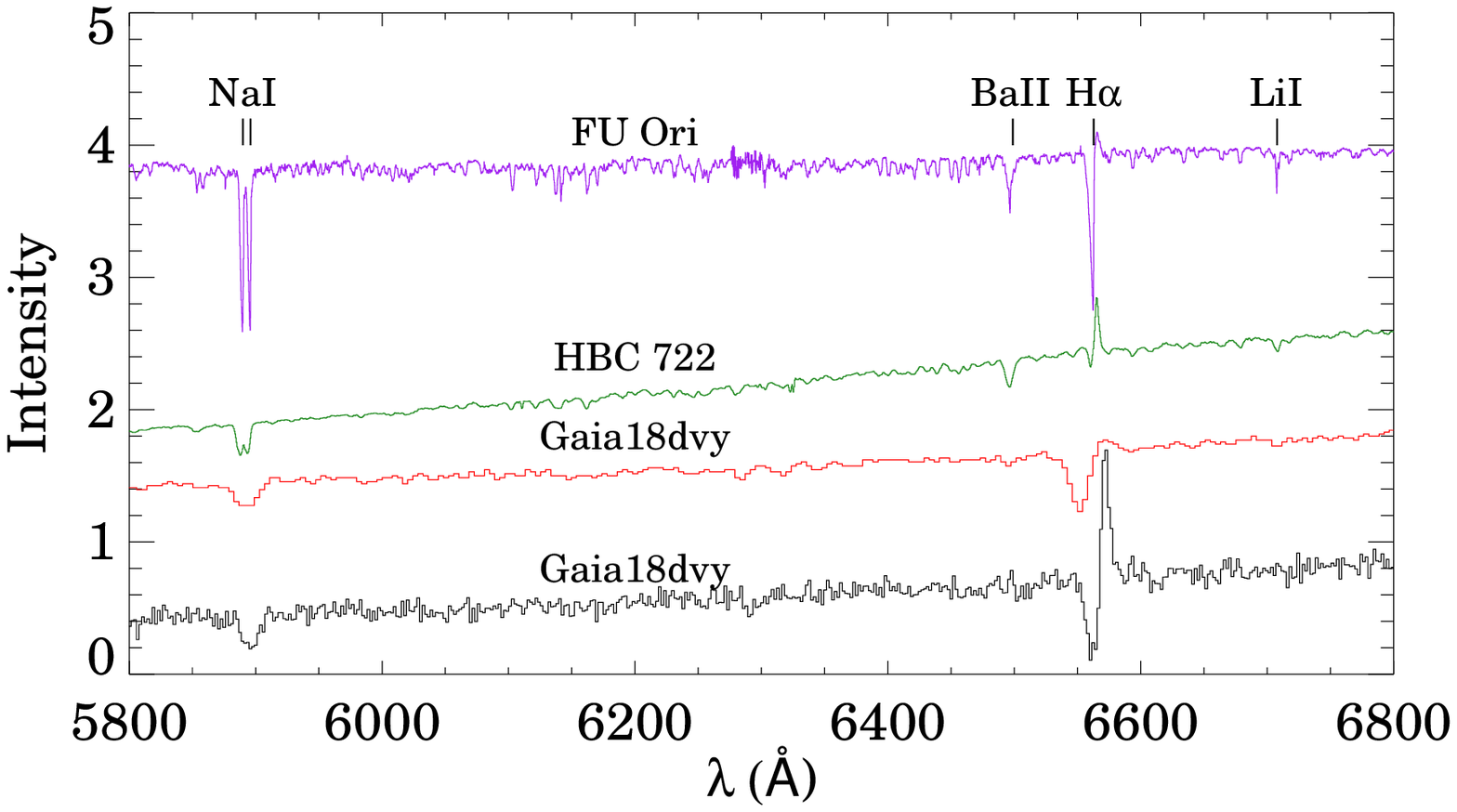}
\includegraphics[width=\columnwidth]{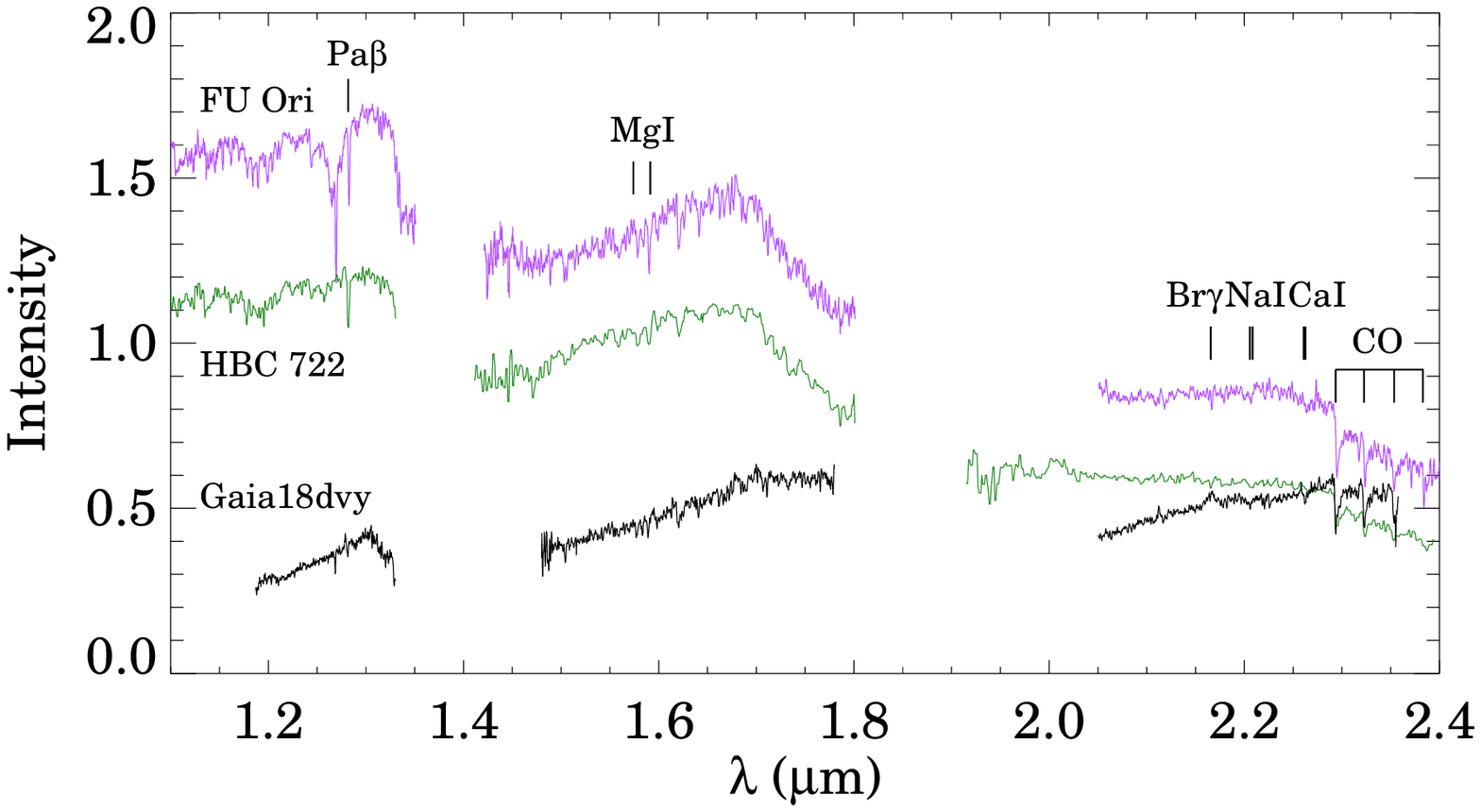}
\caption{Top: Portion of the optical spectra of Gaia\,18dvy compared to a VLT/XSHOOTER spectrum of FU~Ori (ESO archival data from program 094.C-0233), and a GTC/OSIRIS spectrum of HBC~722 \citep{kospal2016}. Bottom: Infrared spectra of Gaia\,18dvy, FU~Ori, and HBC~722. Units are arbitrary.}\label{fig:spec}
\end{figure}

The results of our spectroscopic observations are displayed in Fig.~\ref{fig:spec}.

\section{Results}
\label{sec:res}

\subsection{The distance of Gaia\,18dvy}

The position of Gaia\,18dvy is projected on the west periphery of the Cygnus~OB3 association. The star's {\gaia}-based distance, published by \citet{bailer2018}, $4.6_{-1.9}^{+3.3}$\,kpc, is quite uncertain, because the object was faint at the beginning of the {\gaia} mission. To study the relationship between Cygnus~OB3 and Gaia\,18dvy, we compared the {\gaia} DR2 proper motion \citep{gaiadr2} of Gaia\,18dvy with those of bright members of Cygnus~OB3 \citep{tag1,tag2,tag3}, and found good agreement. This suggests that Gaia\,18dvy can be a member of the Cygnus~OB3 association. To estimate the distance of Cygnus~OB3, we plotted the distribution of distances from \citet{bailer2018} for the bright members of Cygnus~OB3, and found a distinct peak at $1.88$~kpc.
We adopt this value as the distance of Gaia\,18dvy.

\subsection{Light curves and color variations}

Pre-outburst photometric observations (IPHAS and Pan-STARRS at optical, 2MASS and UKIDSS in the infrared) imply that Gaia\,18dvy had been faint at least for a decade before 2015. The {\gaia} light curve (Fig.\,\ref{fig:light}) demonstrates that the quiescent phase continued at optical wavelengths until September 2017, when a gradual brightening began. The highest brightening rate was 0.42\,mag/month in the $G$-band. The rapid rise was also documented by ZTF with a similar rate, suggesting an almost wavelength-independent brightening in the optical.

The outburst of Gaia\,18dvy was also seen in the mid-infrared with {\wise} (Fig.\,\ref{fig:light}). Between early 2015 and late 2018 the  brightening at 3.4 (4.6)\,$\micron$ was 1.3 (1.1)\,mag, somewhat lower than the $G$-band rise of 1.6\,mag for the same period. 

Since mid-2019 Gaia\,18dvy is almost constant at all wavelengths, exhibiting a flat maximum. 
The magnitude differences between this maximum and the pre-outburst Pan-STARRS brightness are: ${\Delta}B=4.5$\,mag, ${\Delta}V = 4.3$\,mag, ${\Delta}R\mathrm{\sb{C}} = {\Delta}I\mathrm{\sb{C}}=4.2$\,mag, suggesting that not only the quickest rising phase, but also the whole outburst was almost independent of wavelength, exhibiting only a weak blueing trend as the source became brighter.

The TESS light curve (Fig.~\ref{fig:tess}) outlines stochastic variability with peak-to-peak amplitude of 0.16\,mag, occurring on timescales of 2--3 weeks, and also short-time (several days) events. We calculated the Lomb--Scargle periodogram for two parts of the TESS light curve: before and after its maximum at JD = 2,458,708, after subtracting a linear trend separately for the two parts (the interval 
 JD = 2,458,717 --  2,458,726 was discarded due to a stochastic peak). The results (Fig.~\ref{fig:tess}, bottom) indicate periodic brightness variations in the first part with a period of $P = 2.47{\pm}0.03$\,days that is significant at the 6$\sigma$ level. The double period of $4.86{\pm}0.26$\,d is also observed with even higher significance. While the power spectrum of the second part also shows several peaks (the strongest one at $P = 3.71$\,d) the frequency and power of these peaks depend on whether to include or discard the large stochastic peaks present in this part of the light curve. Extrapolating the $P = 2.47$\,d period to the second part of the light curve turned out to be inconsistent with the data. This suggests that the periodic behavior of Gaia18dvy can change rapidly on a few days time scale. The TESS data samples the flat maximum brightness phase of the outburst. The light curve demonstrates that while the source was relatively stable at this time, smaller scale variability was still present. Similar variability was observed in FU~Ori, and may be due to flickering or inhomogeneities in the accretion disk \citep{kenyon2000,siwak2013}.

The left part of Fig.~\ref{fig:tcd} presents a $V$ vs.~$V-R\mathrm{\sb{C}}$ color-magnitude diagram. The data points suggest that the brightening of the source from the pre-outburst level, represented by the Pan-STARRS average magnitudes before 2014, to the present maximum was almost wavelength-independent. The colors of the brightening are clearly different from the extinction path, marked in the figure, indicating that the outburst was caused by some other mechanism than the removal of obscuring material in the line-of-sight. As we will show in Sec.~\ref{sec:modeling}, this can be attributed to increasing accretion. The data points from 2019 exhibit blueing with increasing $V$ band brightness. This behavior is different from the color changes during the rapid rising part of the outburst, suggesting that the small brightness variations in 2019 were not due to fluctuating accretion. Nor it is caused by variable dust obscuration, as demonstrated by the significantly different slopes of the extinction path and the observations. 

The NIR color$-$color diagram (Fig.~\ref{fig:tcd}, right) shows that in the bright state Gaia\,18dvy seems to be a reddened T~Tauri-type star, whereas in quiescence the NIR colors shift to the right, slightly beyond the area occupied by reddened Class~II young stellar objects. These color changes are very similar to those of the eruptive young star HBC~722 \citep{kospal2011}: the star shifted nearly parallel to the T~Tauri locus, indicating variations in the temperature and/or structure of the inner disk \citep{meyer97}.

\subsection{Spectroscopy}

Our optical spectra (Fig.~\ref{fig:spec}) were taken during the brightening phase. The spectra show gradually rising continuum with the H$\alpha$ line displaying a P~Cygni profile and several distinct absorption features, including the NaI doublet at 5892~$\angstrom$ and 5898~$\angstrom$. The absorption feature at 6497~$\angstrom$, observed in the spectra of several FUors and associated with Ba\,II/Ca\,I/Fe\,I blend, and the youth indicator Li\,I at 6709~$\angstrom$ are also discernible. Except for the different profiles of H$\alpha$, our two spectra of Gaia\,18dvy are very similar. Our NIR spectrum (Fig.~\ref{fig:spec}) shows several distinct spectral features, most of them in absorption. The Paschen~$\beta$ line can be identified with a small P~Cygni profile. The drop of the spectrum around 1.3\,$\micron$ indicates the beginning of a broad water band. We could identify a few metallic lines: Mg\,I at 1.57 and 1.58\,$\micron$, Na\,I at 2.21\,$\micron$, and Ca\,I at 2.26\,$\micron$. The detection of Br$\gamma$ is uncertain. From 2.3\,$\micron$ a very prominent CO bandhead absorption is visible. 


\section{Modeling}
\label{sec:modeling}

To characterize Gaia\,18dvy in the pre-outburst state, we compiled its spectral energy distribution (SED) from photometric measurements obtained before 2015. In the optical, we adopted the average Pan-STARRS magnitudes. In the infrared, we used UKIDSS {\it JHK$_{\rm s}$} and WISE 3.4--22\,$\micron$ photometry. For comparison, we also compiled an SED for the peak brightness in 2019 as well as for two epochs representative of the rapid brightening phase in 2019, using ZTF, WISE, and our own photometry. All four SEDs are plotted in Fig.~\ref{fig:RT:E}.

\subsection{The central star}
\label{sec:photosphere}

We determined the spectral type and line-of-sight extinction of the central star by comparing the observed $B-V$, $V-I_{\rm C}$, and $I_{\rm C}-J$ colors to reddened color indices of pre-main sequence stars from \citet{mamajek2013}, on a grid of $2880\,{\rm K} < T_{\rm eff}< 7280$\,K and $0<A_V<10$\,mag. At each grid point we reddened the intrinsic colors according to the extinction law of \citet{cardelli89} using $R_V = 3.1$ and calculated $\chi^2$. 
Although there is a degeneracy between $T_{\rm eff}$ and $A_V$, we found  two local minima, one at $T_{\rm eff} = 4330$\,K and $A_V = 3$\,mag ($L_{\ast}=0.8  L_{\odot}$), and another at $T_{\rm eff} = 6900$\,K and $A_V = 5.2$\,mag ($L_{\ast}=2.9 L_{\odot}$). A comparison with pre-main sequence evolutionary tracks \citep[e.g.][]{palla2012} suggests that the first minimum corresponds to a few million years old T~Tauri star \citep[spectral type K4,][]{mamajek2013}, while the second one is an F1-type star already on the zero age main sequence. Since Gaia18dvy is still surrounded by a circumstellar disk, and since the known precursors of most FUors are low-mass objects, we will adopt $T_{\rm eff} = 4330$\,K and $A_V = 3$\,mag in the subsequent disk models. This choice is also supported by the fact that its extinction is broadly consistent with the value of $A_V \leq 2$\,mag extracted from the 3D all-sky maps of \citet{green2019}.

\subsection{The quiescent disk}
\label{sec:quiescentdisk}

To describe the geometry of the circumstellar matter in quiescence, we performed {\RT} modeling of the quiescent  SED, using the RADMC3D code \citep{radmc3d}. For the central star we used a \citet{castelli2004} model with $T_{\rm eff}$ and $A_V$ as above. We fixed the surface gravity to $\log g = 3.5$ and metallicity to $m = 0$. For the disk, we assumed power-law density distribution \citep{chen2018}, with inner and outer radii $R_{\rm in}$ and $R_{\rm out}$, surface density power-law index $p$, scale height power-law index $q$, inner dimensionless scale height $h_{\rm in}$, and mid-plane opacity $\tau$. For dust composition, we assumed 1:1 mixture of amorphous carbon and interstellar silicate, and power-law grain size distribution with index of 3.5, from $a_{\rm min}=0.01\,\micron$ to $a_{\rm max}=10^3\,\micron$. 
Fig.~\ref{fig:RT:E} shows our best-fit quiescent model, which has the following parameters:
$L_*=0.8\,L_{\sun}$, $R_{\rm in}=0.2\,$au,
$R_{\rm out}=300\,$au, $h_{\rm in}=0.17$, $p=-1.0$, $q=0.05$,
$i=30\degr$.
The total (gas+dust) mass of the disk is ${\sim}3.9\times10^{-3}\,M_{\odot}$.
The model requires an unusually large inner scale height of $h_\mathrm{in}=0.17$, indicating that, in order to reproduce the measured strong IR excess, a large fraction of stellar light has to be reprocessed by the circumstellar material.
The inner disk radius in the best-fit model is larger than the dust sublimation radius by a factor of ${\sim}5$.
The modeled bolometric luminosity of the system is ${\sim}1.5~L_\sun$.
We note that all these values depend on the luminosity of the central object: adopting a hotter and more luminous star would result in somewhat lower inner scale height. We also caution that the disk mass is poorly constrained with only optical-IR photometry.


\subsection{Accretion disk in the outburst}
\label{sec:accdisk}


In a FUor outburst, the optical--mid-infrared flux is almost exclusively emitted from a hot, luminous accretion disk in the innermost part of the system \citep{hk96}. It can be modeled with a steady, optically thick, geometrically thin viscous gas disk, whose mass accretion rate is constant in radial direction. The inner edge of such a disk is usually set to the stellar radius, while the outer radius is less defined, since it may overlap with the outer cold passive disk. E.g. modeling the FUor V582~Aur with a similar geometry,  \citet{abraham18} adopted 2~au for the outer size of the heavily accreting gas disk (noting that the exact value has no noticeable effect on the results), while the outer cold circumstellar disk extended to much larger radii.

\begin{figure}
\includegraphics[width=\columnwidth]{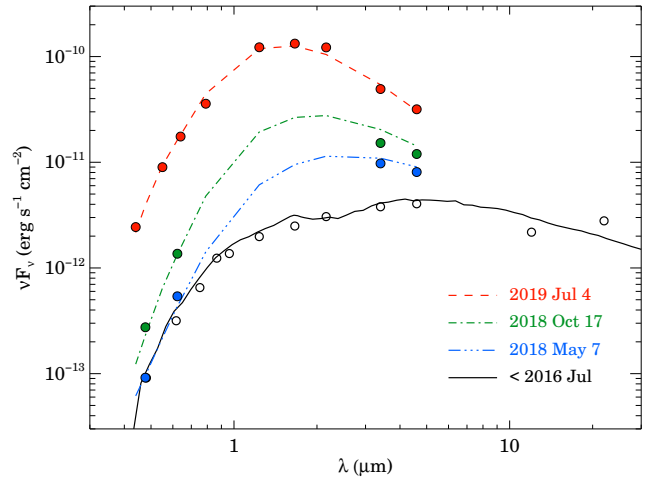}
\caption{Multiepoch SEDs of Gaia\,18dvy: quiescence (black circles), peak of the outburst (red dots), and two epochs representing the brightening phase (blue and green dots). The black curve is our best-fitting RADMC3D model to the quiescent measurements, while the other curves are our best-fitting accretion disk models in excess of the quiescent SED.}
    \label{fig:RT:E}
\end{figure}

To determine the accretion rate and separate the effects of changing extinction and accretion during brightening, we fitted the outburst SEDs (Fig.~\ref{fig:RT:E}) using  the accretion disk model described above. We calculated the disk's flux by summing up the blackbody emission from concentric annuli between the stellar radius and {\bf }$R_{\rm acc}$ following \citet{kospal2016}. We assumed a stellar mass of 1\,$M_{\odot}$, and a disk inclination of 30$\degr$. The stellar radius was computed from the effective temperature and extinction obtained in Sect.~\ref{sec:photosphere}, which resulted in $R_{\rm star}=$ 1.6\,$R_{\odot}$.  It is an unusual feature of the accretion disk modeling of Gaia18dvy that the outer radius, $R_{\rm acc}$, is well constrained by the mid-infrared WISE observations: adopting in a first step $R_{\rm acc}$=2.0~au led to a significant overestimation of the measured mid-infrared fluxes. This result may suggest an unusually small inner accretion disk, and that the outer dust disk has little contribution at these wavelengths. We could reproduce the WISE fluxes by fixing $R_{\rm acc}$ to 0.1\,au. Thus only two free parameters remained: the product of the stellar mass and the accretion rate $M\dot{M}$, and the line-of-sight extinction $A_V$. We obtained the best accretion disk model by $\chi^2$ minimization, and computed formal uncertainties of the fitted parameters with a Monte Carlo approach. 

The most complete coverage of the optical-infrared SED is available for the peak of the outburst (2019 July 4, Fig.~\ref{fig:RT:E}). We could fit it with $\dot{M} = 6.9{\pm}2.1 \times 10^{-6}\,M_{\odot}$yr$^{-1}$, $A_V = 4.35\pm0.4$\,mag, with a reduced ${\chi}^2$ of 1.3. Figure~\ref{fig:RT:E} shows our best fit model (red curve). The derived extinction value is somewhat higher than what we obtained from the photospheric modeling.
The luminosity of the accretion disk is $\sim 175\,L_{\odot}$. We note that adopting a central star with higher $T_{\rm eff}$ would imply a smaller stellar radius, and therefore a smaller inner radius for the disk, and would require the combination of higher luminosity and larger extinction in the best fit accretion disk model.

In a second step, we modeled several additional epochs, where mid-infrared photometric points from WISE and an interpolated G-band magnitude from Gaia were available. We fitted these SEDs by fixing the extinction to the value determined at the peak epoch ($A_V$ = 4.35 mag) and varied only the accretion rate. This procedure  resulted in reasonable fits. The computed accretion rate values are plotted as a function of time in Fig.~\ref{fig:color}a.

\section{Discussion}
\label{sec:discussion}

\citet{connelley2018} suggested eight distinctive spectroscopic features for FUors. Out of these, Gaia\,18dvy exhibits five: (1) strong CO bandhead absorption in the $K$ band; (2) the shape of the $H$-band spectrum is ``triangular'', due to water vapor bands on each end of the $H$-band window; (3) Pa$\beta$ and Br$\gamma$ lines in absorption; (4) only a few emission lines are detectable in the infrared spectra, especially with P~Cygni profiles; and (5) some metallic lines from Na, Mg and Ca are present. Based on these features and the light curve shape, we suggest that Gaia\,18dvy is a new FU~Orionis-type object.

During a period of 1.5 years, the luminosity of Gaia\,18dvy increased from 1.5\,$L_{\odot}$ to 175\,$L_{\odot}$, a factor of more than 100. This outburst luminosity is typical of FUors \citep{audard2014}. The accretion rate is somewhat lower than in most FUors, but is close to the value computed for HBC~722 ($6{\times}10^{-6}\,M_{\odot}$yr$^{-1}$, \citealt{kospal2016}). The location and displacement of HBC~722 in the NIR color-color diagram (Fig.\,\ref{fig:tcd}) are also similar to those of Gaia\,18dvy.

Our results show that the progenitor of Gaia\,18dvy was a K4-type T~Tauri. Using pre-main sequence evolutionary tracks from \citet{palla1999}, the mass of the star is about 1\,$M_{\odot}$. The star is surrounded by a circumstellar disk whose structure and physical parameters in quiescence are typical of T~Tauri disks. The only unusual parameter is the rather large inner scale height, which is inconsistent with hydrostatic equilibrium (that would be only ${\sim}0.04$ at the inner rim of a T~Tauri disk).

\begin{figure}
   \includegraphics[width=\columnwidth]{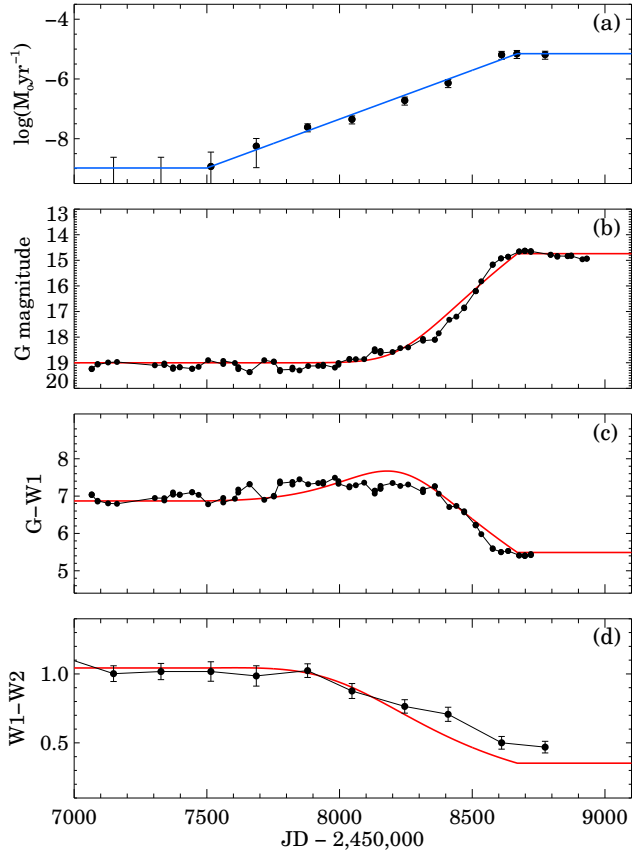}\\
  \caption{{\it (a)}: Optical Gaia light curve of Gaia\,18dvy. {\it (b)}: 
 Optical-infrared color evolution, computed from the Gaia G-band and the {\wise} Band1 magnitudes. {\it (c)}: Mid-infrared color evolution derived from the two {\wise} bands. {\it (d)}: Accretion rates as computed in Sect.~\ref{sec:accdisk}. A simple linear model to the data points is overplotted in blue. Magnitudes and colors, computed from our accretion disk model using \.M values as predicted by the linear model, are overdrawn in the upper three panels.
 }\label{fig:color}
\end{figure}

During the outburst phase, we fitted the observed optical-infared light curves using a simple accretion disk model (Sect.~\ref{sec:accdisk}). Most data points could be reasonably well reproduced by a sequence of models where both the line-of-sight extinction and the disk geometry were fixed, and only the accretion rate was fitted. Figure~\ref{fig:color} summarizes our results. The top panel shows the time evolution of the derived accretion rates, which can be fitted by an exponential function starting at some low values at $<10^{-9}$~M$_{\odot}$yr$^{-1}$ and reaching ${\sim}10^{-5}$~M$_{\odot}$yr$^{-1}$ at the peak of the outburst in mid-2019. Adopting this exponential function (blue lines in Fig.~\ref{fig:color}a) to predict the accretion rate at any given epoch, we computed the various magnitudes and colors as a function of time from the accretion disk model. These results are overplotted in Fig.~\ref{fig:color} (b--d).

The good match at both optical and infrared wavelengths imply that the photometric observations preceding the peak brightness can be explained by a simple accretion disk model of exponentially increasing accretion rate. At early phases of the outburst the accretion rate was low, thus the accretion disk had a low temperature and contributed only to the mid-IR part of the SED, but not to the optical. Later, the rising accretion rate led to higher disk temperatures, and the optical fluxes started growing rapidly, causing increasingly bluer G--W1 colors after JD $\sim$2,458,400.

The observed exponential growth of the accretion rate that started already more than 3 years before the brightness peak (Fig.~\ref{fig:color}a) may provide an important constraint on outburst physics. We calculated the e-folding time of the increase, and adopted the resulting $\sim$145 days as an estimate of the dynamical timescale of the outburst. Interpreting it as a Keplerian period, it would correspond to r$\sim$0.54 au. The geometry of our 
accretion disk, however, implies that the outburst is confined to a smaller area than this, to the innermost 0.1 au of the system. This result should be taken into account in outburst model calculations.


Finally we mention a similarity between Gaia18dvy  and the young eruptive star HBC~722. Plotting the V-band light curve of HBC~722 over the Gaia light curve of Gaia18dvy outlines very similar shapes, but the timescale of the HBC~722 light curve is three times shorter, i.e., all changes happened three times faster. We speculate that the brightening of HBC~722 was also caused by an exponential rise of the accretion rate, but with shorter e-folding time. If true, then possibly the same physical mechanism was responsible for both outbursts, suggesting the existence of a general process whose timescale may change from object to object.



\acknowledgements

We thank an anonymous referee, whose questions and comments significantly improved the paper. This project has received funding from the European Research Council (ERC) under the European Union's Horizon 2020 research and innovation programme under grant agreement No 716155 (SACCRED), Lend\"ulet LP2018-7/2019 and KEP-7/2018 of the Hungarian Academy of Sciences, GINOP 2.3.2-15-2016-00003 and PD-128360 of the Hungarian National Research, Development and Innovation Office, Polish NCN DAINA grant 2017/27/L/ST9/03221, European Commission's Horizon 2020 OPTICON grant 730890, Polish MNiSW grant DIR/WK/2018/12, grant No.~S-LL-19-2 of the Research Council of Lithuania, Project No.~176011 ``Dynamics and kinematics of celestial bodies and systems'' of the Ministry of Education, Science and Technological Development of the Republic of Serbia, DFG priority program SPP 1992 ``Exploring the Diversity of Extrasolar Planets'' (WA 1047/11-1), the MINECO (Spanish Ministry of Economy) through grant RTI2018-095076-B-C21 (MINECO/FEDER, UE). The Joan Oró Telescope (TJO) of the Montsec Astronomical Observatory (OAdM) is owned by the Catalan Government and is operated by the Institute for Space Studies of Catalonia (IEEC). MG is supported by the Polish NCN MAESTRO grant 2014/14/A/ST9/00121. We acknowledge ESA {\gaia}, DPAC, and the Photometric Science Alerts Team. We thank Christina Conner, Megan Davis, Alessandro Dellarovere, Hannah Gallamore, Mira Ghazali, Aaron Kruskie, Dylan Mankel, Jesse Leahy--McGregor, Brandon McIntyre, Barrett Ross, Courtney Wicklund, and Evan Zobel for observing Gaia\,18dvy at the Michigan State University Observatory. Based on observations made with the Nordic Optical Telescope, operated by the Nordic Optical Telescope Scientific Association at the Observatorio del Roque de los Muchachos, La Palma, Spain, of the Instituto de Astrof\'\i{}sica de Canarias.
Based on observations obtained with telescopes of the University
Observatory Jena, which is operated by the Astrophysical Institute of the Friedrich-Schiller-University.

\facilities{Gaia, PS1, TESS, NOT, NEOWISE, Asiago:Copernico, ING:Newton, VLT:Kueyen}

\bibliography{paper}{}

\end{document}